\documentclass[11pt]{article}

\usepackage[margin=1in]{geometry}
\usepackage{amsmath,amssymb,amsfonts}
\usepackage{graphicx}
\usepackage{booktabs}
\usepackage{natbib}
\usepackage{hyperref}
\usepackage{xcolor}
\usepackage{enumitem}
\usepackage{bm}
\usepackage{float}
\usepackage{lineno} 

\newcommand{\vect}[1]{\bm{#1}}

\DeclareMathOperator{\softplus}{softplus}
\DeclareMathOperator{\sigmoid}{sigmoid}
\DeclareMathOperator{\logit}{logit}

\title{Six-hourly Probabilistic Precipitation Forecasts Created from Hourly Probabilities}

\author{Thomas M.\ Hamill\thanks{The Weather Company, Atlanta, Georgia; e-mail:
\texttt{tom.hamill@weather.com}}}

\date{\today}

\begin{document}

\maketitle

\begin{abstract} A lightweight neural network is described that produces six-hourly, spatially
detailed probabilistic quantitative precipitation forecasts (PQPF) from a sequence of hourly
spatially detailed probabilistic forecasts, for lead times from 6 to 72\,h. The hourly forecasts
were generated by a ``zero-inflated'' Attention Residual U-Net that outputs the parameters of a
probability distribution including a probability of no precipitation and a weighted mixture of
two Gamma distributions at each grid point for each hourly forecast lead time. The six-hourly
network passes the six consecutive hourly parameter sets, together with summary statistics of the
spatial texture of the raw deterministic forecast, through a small recurrent (GRU) layer and a
multi-layer perceptron, and maps them to a corresponding zero-inflated mixture of three Gamma
distributions describing the six-hourly accumulated precipitation. Training uses sets of hourly
probabilistic forecasts, drawn preferentially from wet and heavy-precipitation cases (importance
sampling), and matched Multi-Radar Multi-Sensor (MRMS) six-hourly precipitation accumulations,
subject to a quality threshold. The resulting six-hourly probabilistic forecasts are well
calibrated and skillful, as demonstrated by reliability diagrams, performance diagrams and Brier
Skill Scores evaluated over an out-of-sample period spanning day 12 through the end of every
month of 2025. The Brier Skill Scores are consistently and significantly higher than those of a
baseline that sums independent draws from the hourly distributions. Forecasts show realistic
modulation of precipitation related to terrain variations. \end{abstract}

\section{Introduction} \label{sec:intro}

There is now a substantial body of literature on methods for the statistical postprocessing of
ensemble forecasts of precipitation for accumulation periods of 6, 12, or 24 hours. See, for
example, \citet{hamill2023nbm} and \citet{jesberg2023}, who describe methods and results for
improving National Blend of Models six-hourly probabilistic precipitation forecasts using long
time series of reforecasts and precipitation reanalyses. Other past examples include
\citet{sloughter2007}, \citet{voisin2010}, \citet{bentzien2012}, \citet{bouballegue2013},
\citet{scheuerer2014}, \citet{messner2014}, \citet{scheuerer2015}, \citet{baran2016},
\citet{zhang2017}, \citet{hamill2017nbm}, \citet{hamill2018}, \citet{gascon2019},
\citet{ghazvinian2020}, and \citet{ghazvinian2022}.

In a previous article, a method was described for generating hourly high-resolution
probabilistic precipitation forecast guidance from a deterministic weather forecast model
simulation through an Attention Residual U-Net deep learning algorithm \citep{hamill2026a}. These
hourly probabilities may be useful for providing context on precipitation event starting and end
times and intensity. Still, in many situations, a user may also want internally
consistent probabilities over longer periods such as a 6-h window (say, the morning or the
afternoon) or a 12-h window (day, night).

When probability forecasts are issued independently for successive hourly periods, combining them
into a multi-hour accumulation forecast is complicated by the serial correlation of precipitation
events. Consecutive hours are often not meteorologically independent, and treating them as such
may produce poorly calibrated accumulation probabilities. \citet{hughes1979} addressed this type
of problem, combining routine 12-h probability-of-precipitation forecasts from National Weather
Service offices into longer-period estimates. Hughes and Sangster applied empirical
joint-probability adjustments for forecast-period dependence and found that warm-season events
behave more nearly independently than cold-season ones. \citet{wilks1990} examined the
combination of forecast probabilities for consecutive precipitation periods more generally and
showed that the joint distribution of successive events, not simply the product of marginal
probabilities, must be accounted for to achieve reliable aggregated guidance. \citet{wilks1991}
further showed how such serial correlation of meteorological events and forecasts can be
represented explicitly within dynamic decision-analytic models, reinforcing that treating
consecutive periods as independent is a simplification with real decision-relevant consequences.

Papers such as these illustrate the problem of serial correlation but do not provide an
actionable algorithm for generating six-hourly probabilities from hourly data. We thus seek an
alternative, inspired by recent success with deep learning models for precipitation statistical
postprocessing \citep[e.g.,][]{hu2023}. The approach to be described trains a model to aggregate
the distributional information carried by six consecutive hourly forecasts into a single
six-hourly predictive distribution. In principle, the same approach could be used for other
accumulation periods such as 3 h, 12 h, or 24 h, though they will not be demonstrated here.

Why a learned network, rather than a simpler adjustment? Several simpler alternatives exist.
Summing independent draws from the hourly distributions (the baseline of Sec.~\ref{sec:baseline})
ignores serial correlation entirely, and its overconfidence is documented in
Sec.~\ref{sec:results}. One could instead assume a single fixed correlation between hours, but
the dependence between consecutive hours varies with season, regime and the spatial character of
the forecast precipitation \citep{hughes1979,wilks1990}. One could also recalibrate the
baseline's output with a monotone mapping of forecast probability to observed relative frequency,
such as isotonic regression. We did not pursue this for three reasons. First, such a mapping is a
function of the baseline probability alone, fitted separately for each threshold, so it cannot
use any information about how the hours relate (the spatial texture, cross-hour coherence, season
and terrain predictors), and the recalibrated probabilities at different thresholds need not form
a single coherent distribution. Second, a monotone recalibration of an overconfident forecast
necessarily pulls its highest probabilities toward the observed frequencies. The baseline's 90\%
bin, for example, verifies only about 77\% of the time at 0.25\,mm and 12\,h
(Sec.~\ref{sec:results}), so a recalibrated baseline could rarely justify probabilities much
above that value. High-end probabilities are valuable to users, and the network issues them where
warranted: its 90\% bin verifies about 90\% of the time. Third, the network predicts the full
distribution, so all thresholds, including those not used in training or verification, are
available from one consistent set of parameters. We therefore treat improved calibration as
arising from better modeling of the temporal dependence, not from a posteriori adjustment.

This article describes the trained model, a lightweight neural network that maps a
73-dimensional feature vector to the ten parameters of a six-hourly zero-inflated three-component
Gamma mixture, i.e., a probability mass at zero and a weighted three-component Gamma mixture model
for positive precipitation amount. The feature vector combines six consecutive hourly
gamma-mixture parameter sets with seasonal, time-of-day and terrain predictors and with
statistics of the spatial texture and cross-hour coherence of the raw deterministic forecast. The
six hourly parameter sets (plus three texture statistics per hour) are first summarized by a
small gated recurrent unit \citep[GRU;][]{cho2014}, whose output is combined with the remaining
features in a fully connected network \citep[MLP;][]{rumelhart1986}. The network is trained
separately for each lead time, from 6 to 72\,h in 6-h steps, using sets of six hourly probabilistic
forecasts and matched MRMS six-hourly precipitation sums, sampled preferentially from wet and
heavy-precipitation cases (importance sampling; Sec.~\ref{sec:sampling}) because most grid points
are dry. By operating on the distributional parameters of the hourly forecasts rather than on raw
precipitation fields, the approach avoids the need to store or process full-resolution spatial
fields at training time, and the resulting six-hourly probabilistic forecasts inherit the
calibration properties of the underlying hourly model.

The remainder of this document is organized as follows. Section~\ref{sec:hourly} describes the
input data, including the probabilistic forecast feature data, the MRMS target precipitation, and
the sampling strategy used to construct the training set. Section~\ref{sec:mlp} describes the
network architecture, loss function, and training. Section~\ref{sec:results} presents Brier Skill
Scores, reliability diagrams and performance diagrams and compares them with a reference that
assumes independence between hours. Section~\ref{sec:summary} provides a summary and possible
next steps.

\section{Input data} \label{sec:hourly}

\subsection{Probabilistic forecast feature data} \label{sec:input_feat}

The hourly probabilistic forecasts used as input are produced by an Attention Residual U-Net
(Attention ResUNet) applied to The Weather Company's GRAF (Global high-Resolution Atmospheric
Forecasting) model output. GRAF is a convection-permitting model based on the NCAR Model for
Predictions Across Scales \citep[MPAS;][]{skamarock2012}, with mesh refinement to approximately
4.25\,km grid spacing over the contiguous United States (CONUS) and Europe. The hourly
postprocessing method was introduced by \citet{hamill2026a}. The hourly forecasts used here come
from the subsequent version of that method described by \citet{hamill2026b}, in which a single
network per season, conditioned on calendar season and forecast lead time through Feature-wise
Linear Modulation, replaces separately trained per-month, per-lead networks, and forecasts extend
to 72\,h. The same hourly model supplies the inputs at all leads from 6 to 72\,h.

\begin{figure}[H] \centering
\includegraphics[width=\textwidth]{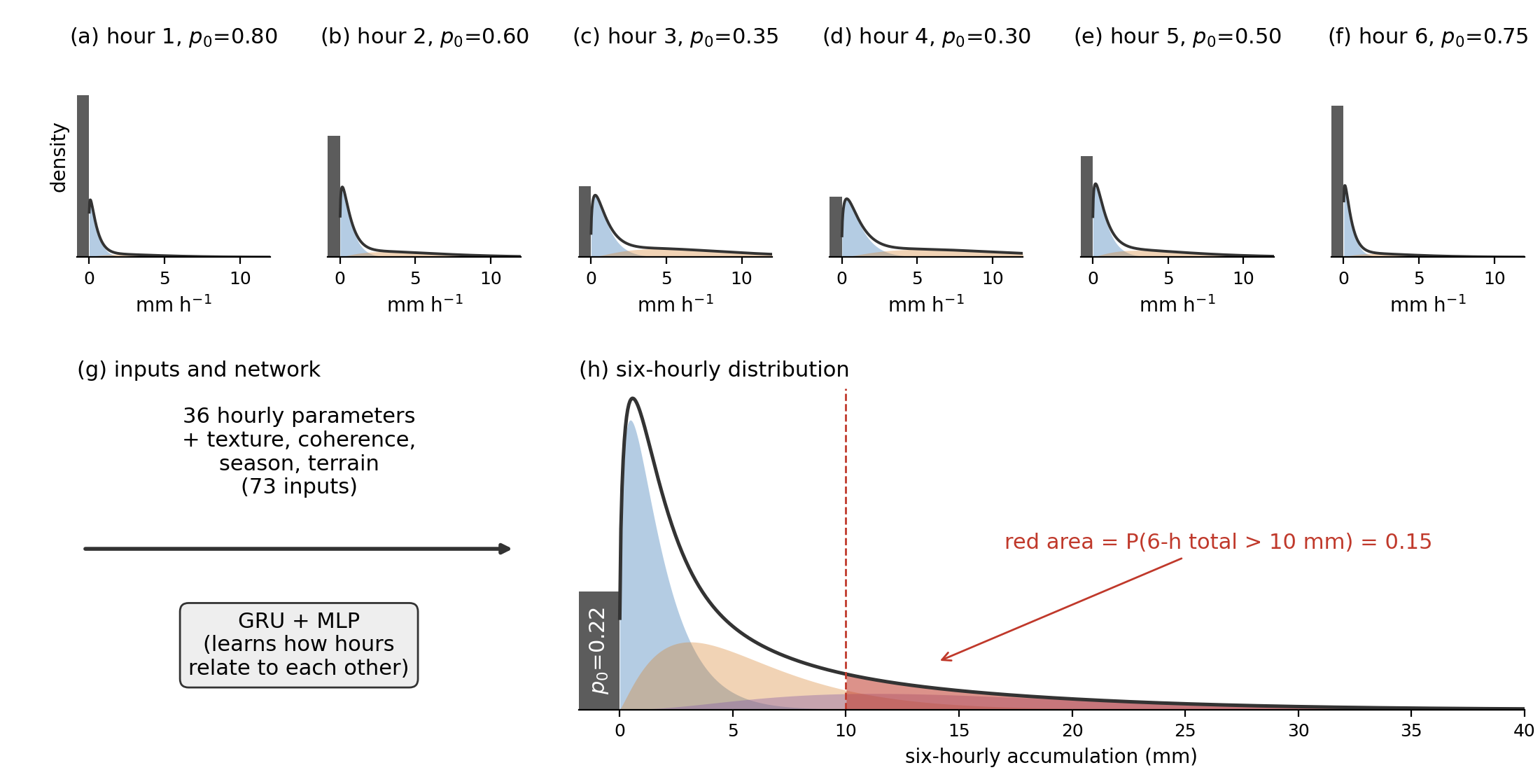} \caption{Schematic of
the aggregation. (a--f) Six hourly zero-inflated Gamma-mixture forecasts at one grid point. In
each panel the dark bar is the probability of no precipitation $p_0$ and the curve is the density
of the wet amount, a weighted sum of two Gamma components (blue, orange). (g) The six hourly
parameter sets, together with the other input features, are passed to the six-hourly network (GRU
and MLP). (h) The network returns one zero-inflated three-component Gamma mixture (blue, orange,
purple components) for the six-hour total; the shaded area above a threshold (here 10\,mm) is the
exceedance probability, and the probability for any other threshold follows from the same
distribution. Parameter values are hypothetical and chosen only to show the structure, not real
forecast output.} \label{fig:schematic} \end{figure}

Precipitation amount at a point is awkward to describe statistically: it is exactly zero much of
the time, and when it is positive it is strongly skewed, with many light amounts and a long tail
of rare heavy amounts. A zero-inflated Gamma mixture handles these two features separately. A
single number, $p_0$, is the probability of exactly zero precipitation. Given that it is wet, the
amount is described by a weighted sum of Gamma distributions. A Gamma distribution is a flexible,
strictly positive, right-skewed probability distribution controlled by a shape and a scale
parameter (small shape gives a sharply peaked curve near zero, larger shape moves the peak to
heavier amounts). Using several Gamma components, each with a weight, lets the mixture represent,
for example, a bulk of light drizzle plus a distinct tail of convective downpours, which one
Gamma curve fits poorly. The probability of exceeding any threshold, say 10\,mm, is
then the probability of nonzero precipitation, $1-p_0$, times the area under the wet-amount curve
above that threshold, which is available in closed form.
The hourly forecasts use two components; the six-hourly forecast uses three because totals over
six hours may experience a greater variety of precipitation phenomena. Figure~\ref{fig:schematic}
illustrates the idea: six individual hourly distributions at one grid point enter the six-hourly
network, which returns a single distribution for the six-hour total. Because the output is a
distribution rather than a number, the probability of exceeding any threshold follows from it
consistently.

At each grid point and for each hourly lead time, the Attention ResUNet predicts the six
parameters of a zero-inflated two-component Gamma mixture: \begin{equation} \vect{\psi}_h = (p_0,\;
w,\; \alpha_1,\; \theta_1,\; \alpha_2,\; \theta_2)_h, \label{eq:hourly_params} \end{equation}
where $p_0$ is the probability of zero hourly precipitation, $w$ is the conditional mixing weight
for the lighter component, and $(\alpha_j, \theta_j)$ are the shape and scale parameters of the
$j$-th Gamma component. Component 1 is constrained to represent lighter precipitation than
component 2.

The probability of exceeding any precipitation amount threshold $\tau > 0$ is obtained
analytically from the mixture cumulative distribution function. The parameters $\vect{\psi}_h$ are
saved for each grid point, initial condition date and lead time, and these files serve as the
primary data source for both the six-hourly training and inference pipelines described below.

The input feature vector for the six-hourly network has 73 elements, summarized in
Table~\ref{tab:features}. These were arrived at through extensive trial and error. The first
36 are the six distributional parameters from each of the six hourly forecasts valid in the
accumulation window, ordered by parameter and then from earliest to latest hour,
\begin{linenomath} \begin{multline} \vect{x}_{\text{hourly}} = \bigl[ p_{0,1},\ldots,p_{0,6},\;
w_1,\ldots,w_6,\; \alpha_{1,1},\ldots,\alpha_{1,6},\; \theta_{1,1},\ldots,\theta_{1,6}, \\
\alpha_{2,1},\ldots,\alpha_{2,6},\; \theta_{2,1},\ldots,\theta_{2,6} \bigr] \in \mathbb{R}^{36}.
\label{eq:feature_vec} \end{multline} \end{linenomath} They are supplemented by:
\begin{itemize}[nosep] \item \emph{Seasonal and time-of-day dependence} (4 features): $\cos(2\pi
d/365)$ and $\sin(2\pi d/365)$, where $d$ is the initial condition's day of year, and $\cos(2\pi
c/24)$, $\sin(2\pi c/24)$, where $c$ is the initial condition's cycle hour (00, 06, 12, or
18\,UTC). Neither can be inferred from the hourly distributional parameters alone. \item
\emph{Terrain roughness} (1 feature): $\log(1+s)$, where $s$ is the local standard deviation of
GRAF terrain elevation after smoothing at a scale of approximately 60\,km. \item \emph{Terrain
structure} (3 features): the local terrain-height difference and the smoothed terrain gradients
in the longitudinal and latitudinal directions, the same static fields used as predictors by the
hourly Attention ResUNet. \item \emph{Spatial texture of the raw forecast} (18 features): for
each of the six hours, three statistics computed from the raw GRAF hourly precipitation in a
$15\times15$ grid-point window (about 64\,km) centered on the point: the fraction of the window
with at least 0.1\,mm (wet-area fraction), the ratio of the window's maximum to its mean
(peak-to-mean ratio), and the window's coefficient of variation. The last two are transformed by
$\log(1+x)$. These statistics describe whether the forecast precipitation is a coherent area or
scattered cells, information that the per-point marginal distributions in
Eq.~\eqref{eq:hourly_params} do not carry. \item \emph{Cross-hour coherence of the raw forecast}
(10 features): for each of the five pairs of consecutive hours, the Jaccard overlap of the wet
masks in the window and the z-scored spatial pattern correlation of the two hourly fields. \item
\emph{Raw six-hour forecast total} (1 feature): the GRAF deterministic six-hourly precipitation,
transformed by $\log(1+x)$. \end{itemize}

\begin{table}[H] \centering \caption{Input features of the six-hourly network. The first two
groups are per-hour quantities fed to the recurrent layer; the remainder are constant over the
window.} \label{tab:features} \begin{tabular}{p{0.66\textwidth}cc} \toprule Feature group &
Number & Treatment \\ \midrule Hourly gamma-mixture parameters
($p_0,w,\alpha_1,\theta_1,\alpha_2,\theta_2$) $\times$ 6\,h & 36 & GRU sequence \\ Texture
statistics $\times$ 6\,h (wet-area fraction, peak-to-mean, coefficient of variation) & 18 & GRU
sequence \\ Cross-hour coherence $\times$ 5 pairs (Jaccard, pattern correlation) & 10 & static \\
Day-of-year and cycle-hour cyclic pairs & 4 & static \\ Terrain roughness, terrain height
difference, terrain gradients & 4 & static \\ Raw GRAF six-hour total & 1 & static \\ \midrule
Total & 73 & \\ \bottomrule \end{tabular} \end{table}

\subsection{Target MRMS data} \label{sec:mrms}

The training target at each sampled grid point is the six-hourly Multi-Radar Multi-Sensor
\citep[MRMS;][]{zhang2016} precipitation accumulation, computed as the sum of the six matched
hourly MRMS analyses: \begin{equation} y = \sum_{k=1}^{6} r_k, \quad y \geq 0 \text{ mm},
\label{eq:target} \end{equation} where $r_k$ is the MRMS hourly precipitation for the $k$-th lead
time.

Grid points are included in the training set only if their mean MRMS data-quality index over the
six verification hours exceeds a threshold $Q = 0.6$. This is more restrictive than the quality
threshold used for training the hourly Attention ResUNet ($Q > 0.01$), because the six-hourly
accumulation compounds any retrieval uncertainty across six hours; a higher quality threshold
reduces the risk of training on accumulations that are substantially influenced by radar-beam
blockage or other retrieval problems. The same threshold is applied in verification.

\subsection{Sampling strategy} \label{sec:sampling}

Training data are assembled from previously generated hourly probability distribution parameters.
Training and verification use the CONUS domain only (the Europe domain mentioned in
Sec.~\ref{sec:input_feat} is not used here). Samples are drawn from the first nine calendar days
of every month of 2025, from forecasts initialized at 00, 06, 12 and 18\,UTC. Spanning all twelve
months and all four daily cycles lets the network learn genuine seasonal and cycle-dependent
variation through the cyclic predictors of Sec.~\ref{sec:input_feat}. Days 10 and 11 of each
month are withheld entirely as a gap, and days 12 through the end of the month are reserved for
out-of-sample testing (Sec.~\ref{sec:results}); the gap prevents a persistent synoptic system
from straddling the training/testing boundary and inflating apparent skill. A separate network is
trained for each lead time; the six hourly parameter sets used are those valid in the six hours
ending at the lead time. For example, hourly forecast parameters for hours 19--24 are used when
training the 24-h forecast of the accumulation over hours 19--24.

Because most grid points are dry at any given time, uniform random sampling would yield a
severely imbalanced training set, and would contain few heavy-precipitation cases. Instead, each
grid point is assigned a sampling weight that increases with both its likelihood of being wet and
the forecast's own expected accumulation. The likelihood of positive precipitation during the
window is estimated from the minimum fraction-zero probability across the six hourly lead times,
$p_{\text{nz}} = 1 - \min_k p_{0,k}$. The expected accumulation $\mu_{6}$ is the sum over the six
hours of the means of the zero-inflated hourly mixtures, $\mu_6 =
\sum_{k=1}^{6}(1-p_{0,k})\,[w_k\alpha_{1,k}\theta_{1,k} + (1-w_k)\alpha_{2,k}\theta_{2,k}]$. The
weight is \begin{equation} p_{\text{samp}} = a + (1-a)\,p_{\text{nz}} + B\,\frac{\mu_6}{\mu_6 +
M}, \label{eq:importance} \end{equation} where $a = 0.01$ ensures all quality-passing grid points
retain a small probability of being selected, $B = 8$ sets the emphasis on heavy forecast
accumulations, and $M = 20$\,mm is the expected accumulation at which the last term reaches half
of its maximum. The minimum (rather than, e.g., one minus the product of the six $p_{0,k}$ values, which
would estimate the probability of at least one wet hour under an independence assumption) is used
for $p_{\text{nz}}$ because it identifies the single most-likely-wet hour in the window directly
from the forecast itself, without invoking the independence assumption that
Sec.~\ref{sec:results} explicitly contrasts the network against. Terrain does not enter the
sampling weights; it enters only as a feature.

Candidate points are drawn without replacement according to the normalized $p_{\text{samp}}$
weights, subject to the MRMS quality criterion, from a candidate pool six times larger than the
target sample size. They are then thinned spatially by accepting candidates in descending weight
order, with at most two accepted points in any $15\times15$ block of grid points (about 64\,km),
until 2500 samples per initial condition are obtained. The thinning prevents many samples from
being drawn from the same storm system. Without it, strongly correlated samples were found to let
the network overfit within the first training epoch. Samples are concatenated over all twelve
months.

\section{Network Architecture, Loss Function, and Training} \label{sec:mlp}

\subsection{Architecture} \label{sec:architecture}

The 73 input features are divided into a sequence part and a static part
(Table~\ref{tab:features}). The sequence consists of six time steps, one per hour of the
accumulation window, each with nine features: the six hourly gamma-mixture parameters and the
three texture statistics of that hour. A single-layer GRU with 48 hidden units reads this
sequence in chronological order, and its final hidden state summarizes the evolution of the hourly
distributions across the window. This summary is concatenated with the 19 static features, giving
67 inputs to a fully connected network with hidden-layer widths $[128, 144, 72, 36, 12]$, which
ends in a linear layer with ten outputs: \begin{equation} \text{(GRU: } 6\times 9 \to 48\text{)}
\;\oplus\; 19 \;\to\; 128 \to 144 \to 72 \to 36 \to 12 \to 10. \label{eq:mlp_arch} \end{equation}

Each hidden layer
applies a linear transformation followed by batch normalization \citep{ioffe2015}, a ReLU
activation \citep{nair2010}, and dropout \citep{srivastava2014}. The dropout probability is 0.15
in all hidden layers except the last, where it is zero; dropout immediately before the output
layer was found to systematically under-sharpen the forecast tails, degrading both Brier Skill
Score and calibration. The first hidden layer was widened from 72 to 128 units so that it expands, rather than
compresses, its 67 inputs. The resulting network has 50{,}202 trainable
parameters, which justifies describing the architecture as lightweight: for comparison, the
hourly Attention ResUNet that supplies the input features has on the order of $3 \times 10^7$
parameters, roughly 600 times as many.

The GRU is included because the six-hourly distribution depends not only on what each hour's
forecast looks like but also on how the hours relate to one another: whether a wet period persists
or is intermittent, whether precipitation is building or decaying, and whether the forecast
texture is coherent from hour to hour. A fully connected network that receives the six hourly
parameter sets as a flat vector must discover these temporal relationships separately for each
input position and has no built-in notion of ordering or of adjacency between hours. The cross-hour
coherence features of Sec.~\ref{sec:input_feat} (Jaccard overlap and pattern correlation between
consecutive hours) supply one hand-chosen description of persistence, but they are a prior guess
about which temporal relationships matter. The GRU instead processes the hours one at a time,
carrying a hidden state forward through a learned gating mechanism that decides, at each step,
how much of the earlier information to retain and how much to update. Because the same weights
are applied at every hour, the network can learn a single description of how one hour's
distribution and texture modify the expectation for the next, using far fewer parameters than an
equivalent flat layer. The resulting 48-element summary is a learned, compact representation of
the temporal dependence that the independence baseline of Sec.~\ref{sec:baseline} ignores. The
sequence is short (six steps) and the hidden state small, so the added cost is modest, and the
hand-engineered coherence features are retained as static inputs rather than replaced. In
development experiments at the 12-h lead time, adding the GRU improved reliability at all
thresholds, with Brier Skill Score unchanged or slightly better, relative to the same network
with a flat input.

\begin{figure}[H] \centering \includegraphics[width=\textwidth]{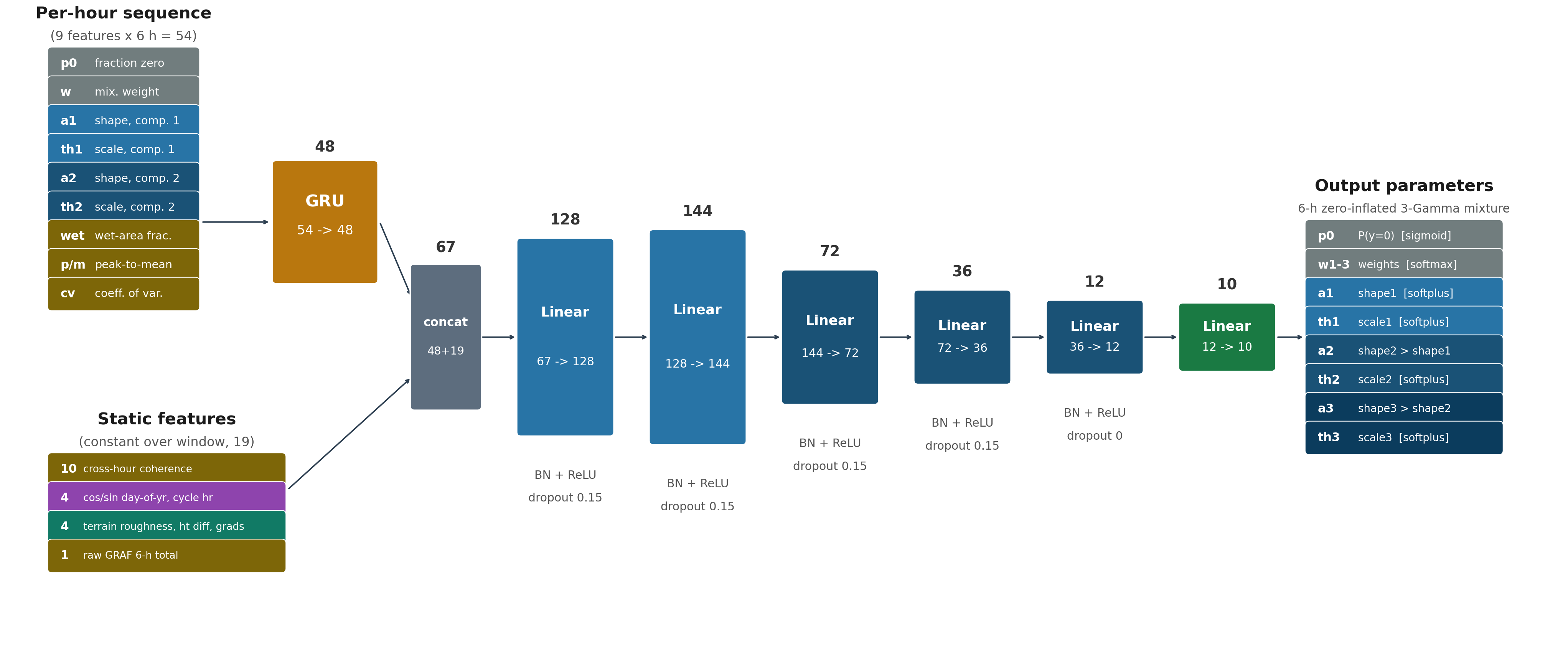}
\caption{Architecture of the neural network for six-hourly probabilistic precipitation
forecasting. The six hourly parameter sets and texture statistics (54 values) are summarized by a
GRU; its 48-element output is concatenated with 19 static features and passed through five hidden
layers ($128 \to 144 \to 72 \to 36 \to 12$, each followed by batch normalization, ReLU and
dropout) to ten outputs. The outputs are transformed via sigmoid (for $p_0$), softmax (mixture
weights) and softplus with a shape-ordering constraint (Gamma shapes and scales) to give the ten
parameters of the six-hourly zero-inflated three-component Gamma mixture.} \label{fig:mlp_arch}
\end{figure}

Figure~\ref{fig:mlp_arch} provides a schematic of the architecture.

\subsection{Output Model Parameters} \label{sec:output}

The ten raw outputs $(z_0,\ldots,z_9)$ are transformed to the ten constrained parameters of the
zero-inflated three-component Gamma mixture $(p_0, w_1, w_2, w_3, \alpha_1, \theta_1, \alpha_2,
\theta_2, \alpha_3, \theta_3)$: \begin{linenomath} \begin{align} p_0 &= \sigmoid(z_0), \\
(w_1,w_2,w_3) &= \mathrm{softmax}(z_1,z_2,z_3), \\ \alpha_1 &= \alpha_{\min} + \softplus(z_4),
\quad \theta_1 = \theta_{\min} + \softplus(z_5), \\ \alpha_2 &= \alpha_1 + \softplus(z_6) +
\Delta_{\min}, \quad \theta_2 = \theta_{\min} + \softplus(z_7), \\ \alpha_3 &= \alpha_2 +
\softplus(z_8) + \Delta_{\min}, \quad \theta_3 = \theta_{\min} + \softplus(z_9), \end{align}
\end{linenomath} where $\alpha_{\min} = 0.1$, $\theta_{\min} = 0.01$, and $\Delta_{\min} = 0.5$.
To resolve label switching among the Gamma components, each component's shape parameter is
reparameterized as a hard structural offset from that of the preceding component rather than by
sorting the components after the fact. This guarantees $\alpha_1 < \alpha_2 < \alpha_3$ by
construction for every input, so no post-hoc reordering step -- and therefore no discontinuous,
non-differentiable relabeling operation -- is needed during training. This extends the
two-component hard-ordering constraint used in the hourly Attention ResUNet's mixture loss
function. A third component gives the distribution more freedom in the heavy-precipitation tail
than the two components used for the hourly forecasts, because the six-hourly totals have a wider
dynamic range.

\subsection{Loss Function} \label{sec:loss}

Training minimizes the negative log-likelihood (NLL) of the zero-inflated three-component Gamma
mixture, in the same form as the loss used for the hourly Attention ResUNet but with an
additional component. For a sample with target six-hourly accumulation $y$:

\paragraph{Case $y = 0$:} \begin{equation} \text{NLL} = -\log(p_0 + \epsilon), \end{equation}
where $\epsilon = 10^{-7}$ avoids taking the logarithm of zero.

\paragraph{Case $y > 0$:} \begin{equation} \text{NLL} = -\log(1 - p_0 + \epsilon) -
\log\!\Bigl[\,\sum_{j=1}^{3} w_j\, f_{\Gamma}(y;\alpha_j,\theta_j)\Bigr], \end{equation}
evaluated in log-space using the log-sum-exp trick for numerical stability. Per-sample NLL values
(a unitless loss, not a precipitation amount) are clamped to a maximum value of 100 before
averaging across the mini-batch, preventing single extreme observations from dominating the
gradient.

The NLL was preferred over the Continuous Ranked Probability Score (CRPS), the other common
strictly proper scoring rule used in probabilistic post-processing. For a zero-inflated Gamma
mixture, the CRPS lacks a simple closed-form expression and must be approximated by numerical integration
or Monte Carlo sampling of the mixture CDF at many threshold values---an expensive inner loop for
every sample in every mini-batch. In contrast, the Gamma log-PDF has an analytic form (computed
via the log-gamma function, \texttt{torch.lgamma}, in PyTorch), so the NLL can be evaluated exactly in a single vectorized pass.
Minimizing the NLL is also equivalent to maximum-likelihood estimation of the distributional
parameters, providing a direct link to the parametric model being fitted.

\subsection{Feature Normalization} \label{sec:normalization}

Each of the 73 input features is independently standardized to zero mean and unit variance using
statistics computed from the training split: \begin{equation} \tilde{x}_k = \frac{x_k -
\mu_k}{\sigma_k}, \label{eq:zscore} \end{equation} where $\mu_k$ and $\sigma_k$ are the
training-set mean and standard deviation of feature $k$. Features with $\sigma_k < 10^{-8}$ are
left unchanged (division by 1 rather than by near-zero). The normalization statistics are saved
and applied consistently at inference time.

Z-score normalization is preferred over the fixed-maximum normalization used for the spatial
Attention ResUNet features because the distributional parameters ($\alpha, \theta$) span a wide range
across lead times and seasons, and their natural scale is not known \textit{a priori}.

\subsection{Climatological Output-Layer Initialization} \label{sec:initialization}

The bias of the final linear layer is initialized using climatological statistics derived from
the training target values, so that the network's first forward pass produces predictions near
the marginal climatological distribution rather than the near-uniform predictions that result
from default (near-zero) weight initialization.

The climatological zero fraction $\hat{p}_0$ is estimated directly from the training targets, and
the bias $b_0 = \logit(\hat{p}_0)$ is set accordingly. A three-component Gamma mixture is then
fitted to the positive training targets (a random subsample of at most 50{,}000 values) by the
expectation-maximization (EM) algorithm, and the biases for the mixture weights, shapes and
scales are initialized to the fitted values via the logarithm of the weights and the inverse
softplus. If too few wet targets are available, a fixed light/medium/heavy set of starting values
is used instead.

\subsection{Training Details} \label{sec:training}

Training is performed with the Adam optimizer \citep{kingma2015} at a learning rate of $\eta =
2.5\times10^{-4}$ and an $L_2$ weight decay of $10^{-4}$. Mini-batches of 1024 samples are
processed. Within the day 1--9 training window of each month, training and validation sets are
split by calendar day rather than by a random shuffle of pooled samples: day 5 of each month is
held out for validation, with all four initialization cycles of that day assigned to the same
split. This day-block scheme yields a more independent validation set than a random split, since
nearby samples in time are strongly correlated synoptically, at the cost of a small validation
set (one day per month) whose NLL estimate is correspondingly noisier than a larger random split
would give; this trade-off was not evaluated quantitatively. A \texttt{ReduceLROnPlateau}
scheduler (reduction factor 0.5, patience 5 epochs) adapts the learning rate based on validation
NLL, and training halts after 5 consecutive epochs without improvement (early stopping), up to a
maximum of 75 epochs. Gradients are clipped to a maximum $\ell_2$ norm of 5.0. The random seed
for weight initialization and mini-batch order is 999 at all leads except 72\,h, where it is
2026; see Sec.~\ref{sec:seed}.

\subsection{Baseline (independent draws)} \label{sec:baseline}

To provide a reference against which the network's aggregation skill can be judged, results also
include a naive \emph{baseline (independent draws)}, labeled ``Independent draws'' in the figures
and referred to in the text as the independence-assumption control. For every verification case,
a 50-member Monte Carlo ensemble is drawn at each grid point from the zero-inflated two-component
Gamma-mixture forecast valid at each of the six hours spanning the accumulation window.
Corresponding ensemble members are then summed across the six hours under the (deliberately
naive) assumption that hourly precipitation amounts are statistically independent from hour to
hour. Six-hourly exceedance probability at each threshold is estimated as the ensemble relative
frequency of the summed accumulation exceeding that threshold. Because the baseline consumes
exactly the same hourly marginal distributions that are fed to the network as input features,
differences between the baseline and the network isolate the effect of the temporal correlation
structure learned from training data, rather than differences in the underlying hourly forecast
quality.

\section{Results} \label{sec:results}

Forecasts are evaluated for each synoptic-cycle (00, 06, 12, 18\,UTC) initial condition from day
12 through the end of every month of 2025 -- an out-of-sample test period disjoint from the day
1--9 training window used in Sec.~\ref{sec:sampling}, comprising 228 calendar days. The
probability of the six-hourly accumulation exceeding a threshold $\tau$ is computed analytically
from the mixture survival function using the regularized incomplete gamma function. Six matched
hourly MRMS files are summed to form the verification six-hourly accumulation, and only grid
points with mean MRMS quality exceeding 0.6 are included in verification, the same threshold used
for the training sample (Sec.~\ref{sec:mrms}). Skill is measured with the Brier Skill Score
(BSS), with the sample climatological frequency of the verification points as the reference.

The network is compared with the baseline (independent draws) described in
Sec.~\ref{sec:baseline}.

\subsection{Brier Skill Score versus lead time} \label{sec:seed}

Uncertainty in the BSS and in the BSS difference between the network and the control is estimated
with a paired bootstrap. The resampling unit is the calendar day, with all four initialization
cycles of a day resampled together because cycles within a day are strongly correlated. The same
resampled days are used for the network and the control, so the confidence interval of the BSS
difference is paired. Collecting spatial samples into a per-case unit before resampling follows
the reasoning of \citet{hamill1999}: nearby grid points within one synoptic event are strongly
correlated, and treating them as independent would overstate the effective sample size. There are
2000 bootstrap replicates.

\begin{figure}[H] \centering
\includegraphics[width=0.55\textwidth]{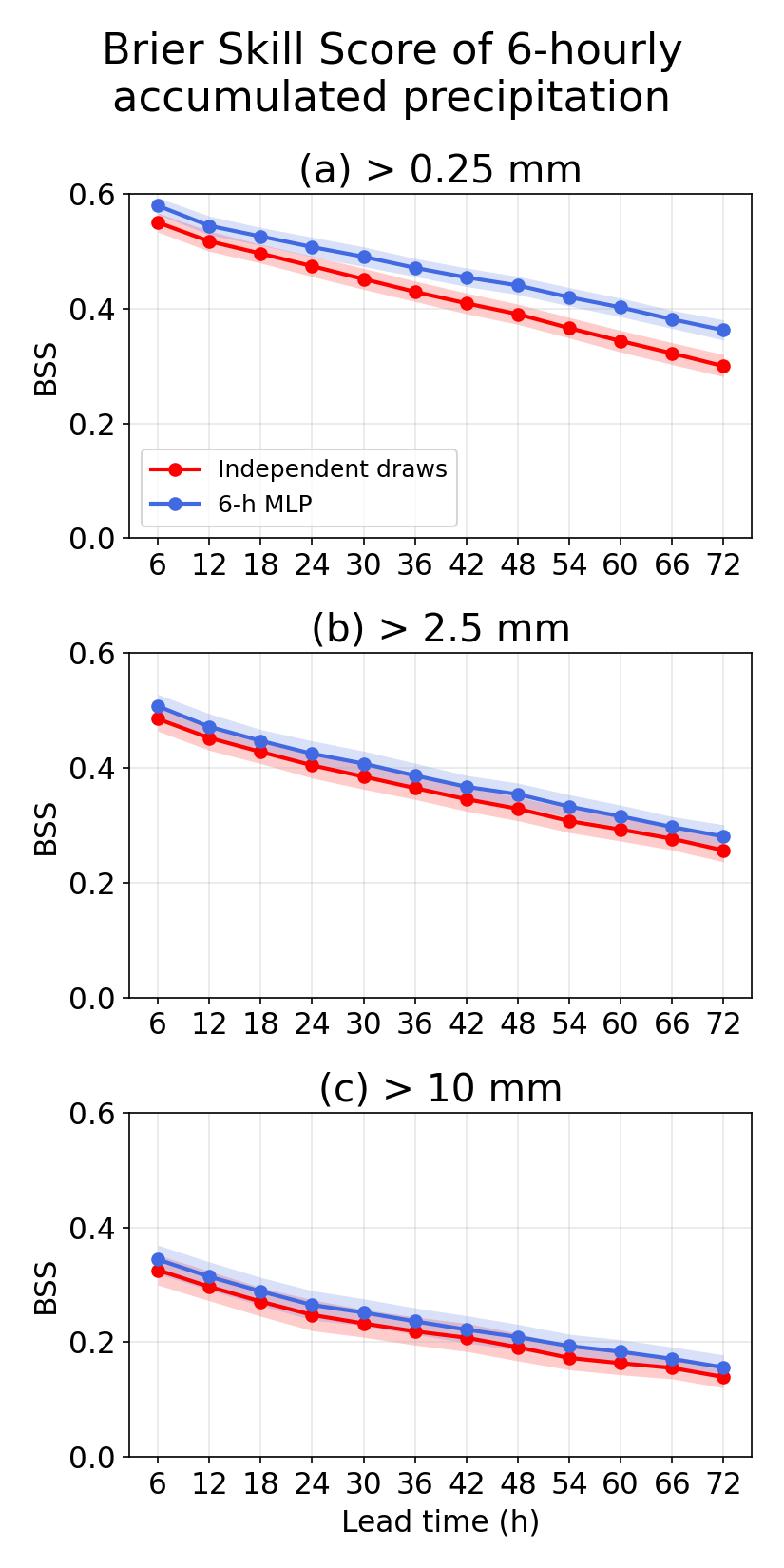} \caption{Brier Skill Score
of six-hourly accumulated precipitation as a function of lead time for the six-hourly network
(blue) and the independence-assumption control (red), for exceedance of (a) 0.25\,mm, (b)
2.5\,mm, and (c) 10\,mm, evaluated over day 12 through the end of every month of 2025. Shading
shows 95\% confidence intervals from a calendar-day bootstrap. All leads use the seed-999 network
except 72\,h, which uses seed 2026 (Sec.~\ref{sec:seed}).} \label{fig:bss_lead} \end{figure}

Figure~\ref{fig:bss_lead} shows the BSS at 6-h to 72-h lead times for thresholds of 0.25, 2.5 and
10\,mm per six hours. The network's BSS exceeds the control's at all 12 lead times and all three
thresholds, and in every case the 95\% bootstrap confidence interval of the difference excludes
zero. The BSS declines with lead time for both methods, as expected. The margin at the 2.5 and
10\,mm thresholds is roughly constant with lead time, at 0.02 and 0.015--0.02, respectively. At
the 0.25\,mm threshold, the margin increases from 0.029 at 6\,h to 0.063 at 72\,h, so that the
network retains more of its skill at long lead times than the control does.

Network weights depend on the random seed used for initialization and mini-batch ordering. Seed
999 is used at all leads from 6 to 66\,h. At 72\,h, however, the seed-999 network proved to be an
unusually poor draw: its 10\,mm BSS only tied that of the control, and it was overconfident at
the heavier forecast probabilities, even though its training and validation loss were
indistinguishable from those of other seeds. Training five networks at 72\,h with different seeds
gave BSS values for the 0.25, 2.5 and 10\,mm thresholds that varied by 0.021, 0.017 and 0.019
between the weakest and strongest seed, a range comparable to the width of a single bootstrap
confidence interval. Seed 999 was the weakest, and seed 2026 was among the strongest. The 72-h
results in this article (Fig.~\ref{fig:bss_lead} and the 72-h reliability diagram) use seed 2026.
They should therefore be regarded as somewhat optimistic relative to a typical seed. Whether seed
999 is also somewhat pessimistic at other leads was not tested.

\subsection{Reliability and performance diagrams}

\begin{figure}[H] \centering \includegraphics[width=\textwidth]{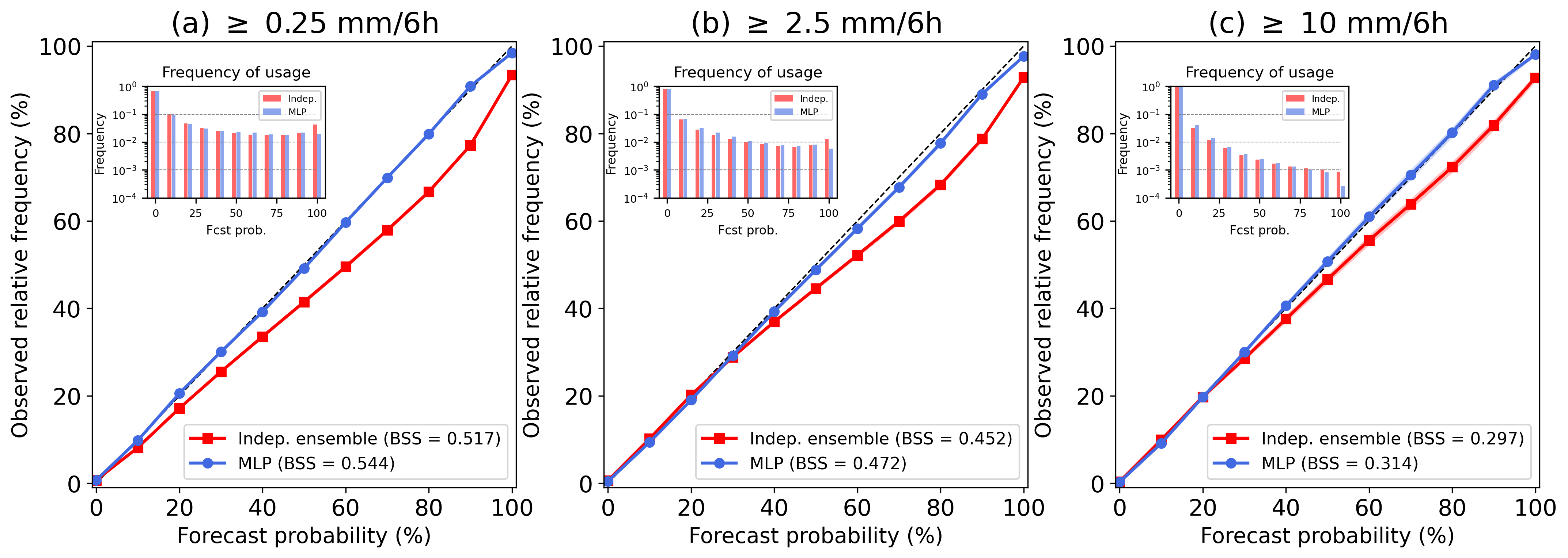}
\caption{Reliability diagrams for six-hourly probabilistic forecasts from the network (blue
circles) at 12-h lead time, evaluated over day 12 through the end of every month of 2025 (an
out-of-sample test period disjoint from the day 1--9 training window; see
Sec.~\ref{sec:sampling}), alongside the naive independence-assumption ensemble control (red
squares; see text). Panels show the reliability of six-hourly exceedance probability for (a)
0.25\,mm, (b) 2.5\,mm, and (c) 10.0\,mm. The diagonal dashed line indicates perfect reliability.
The inset histogram in each panel shows the frequency of use for each probability bin, for both
the network (blue) and the control (red). Shading around each curve is the 10th to 90th
percentile range from a calendar-day bootstrap (2000 replicates). The intervals are very narrow,
because each bin contains many samples, and are hard to see at this scale, especially at 0.25 and
2.5\,mm.} \label{fig:relia_12h} \end{figure}

\begin{figure}[H] \centering \includegraphics[width=\textwidth]{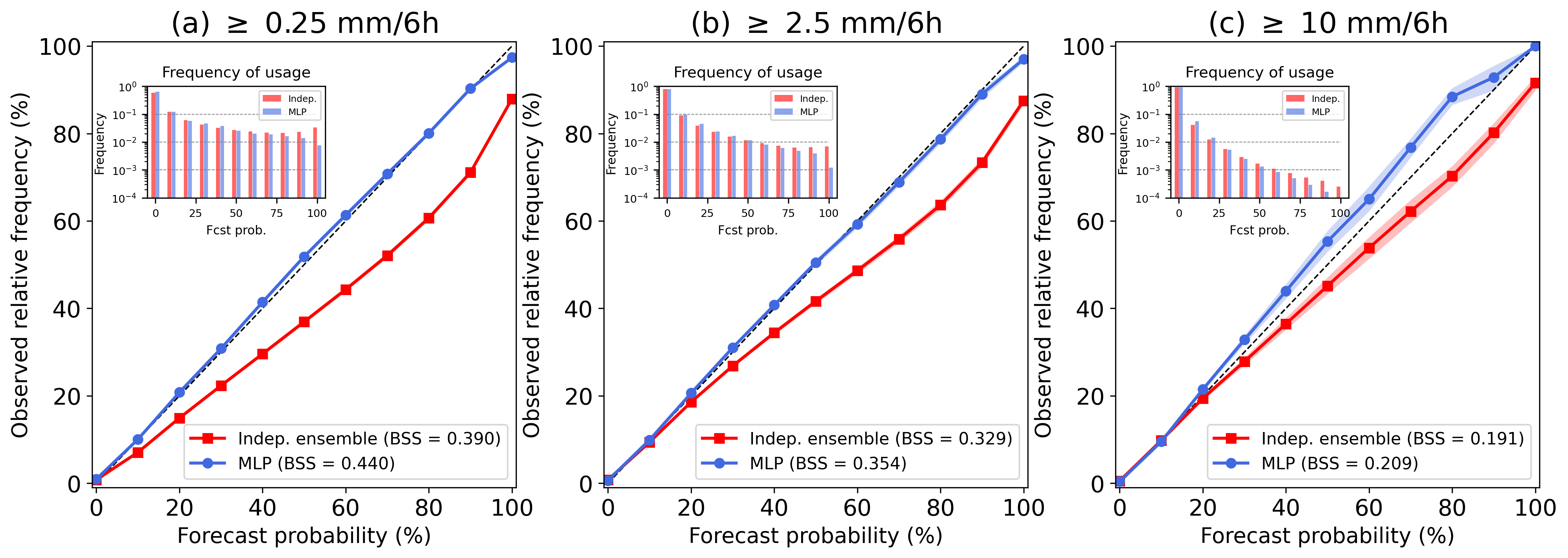}
\caption{As in Fig.~\ref{fig:relia_12h}, but for 48-h lead time.} \label{fig:relia_48h}
\end{figure}

Figures~\ref{fig:relia_12h} and~\ref{fig:relia_48h} show reliability diagrams for the three
thresholds at 12-h and 48-h lead times. Each figure shows the observed frequency of exceedance as
a function of forecast probability for both the network and the independence-assumption control.
The network's forecasts are highly reliable at the 0.25 and 2.5\,mm thresholds at both lead
times, and at the 10\,mm threshold at 12\,h. The independence-assumption control is, by contrast,
consistently overconfident: its reliability curve falls below the diagonal over most of the
probability range, meaning that naively summing independent hourly draws assigns higher
exceedance probabilities than are actually observed. This is consistent with real precipitation
events being temporally autocorrelated -- wet and dry hours often cluster together within a
synoptic event -- so treating the six hourly draws as independent manufactures more small,
partial accumulations than the correlated real-world process produces, inflating the estimated
probability of marginally exceeding a threshold. For example, at the 0.25\,mm threshold and 12-h
lead time, cases assigned a forecast probability of 90\% by the network verify about 90\% of the
time, whereas the control's 90\% bin verifies only about 77\% of the time; the discrepancy is
similar at 48-h lead time (about 90\% for the network vs.\ about 71\% for the control). The
shaded bands in Figs.~\ref{fig:relia_12h} and~\ref{fig:relia_48h} are the 10th to 90th percentile
range of the observed frequency in each probability bin from a calendar-day bootstrap (2000
replicates, resampling days with replacement and summing their daily contingency tables; bins
with five or fewer samples in a replicate are excluded, and a band is drawn only where the bin is
populated in at least half of the replicates). At 0.25 and 2.5\,mm the bands are narrow,
typically a few tenths of a percentage point to about 2 points wide, because the bins are
populated by millions of grid-point samples. The network's curve lies on or within about 1--2
points of the diagonal at nearly all probabilities, and the control's curve is separated from the
diagonal by many times the band width over most of the probability range at 0.25\,mm and by
several times at 2.5\,mm. At 10\,mm and 12\,h the bands widen to roughly $\pm$1 point at the
highest probabilities, and the network's curve remains close to the diagonal.

The 48-h 10\,mm panel (Fig.~\ref{fig:relia_48h}c) is where sampling uncertainty is greatest,
since the number of samples per bin falls from about 2 million at the 30\% bin to about 59{,}000
in the 90\% bin and only 276 in the 100\% bin. All 276 of the latter verified, so every bootstrap
replicate containing them gives an observed frequency of 100\% and the band there has zero width.
The bands are correspondingly wider, about 3--6 points from the 40\% bin upward. However,
sampling uncertainty does not by itself explain the network's underconfidence there. In the 40
through 80\% bins the observed frequency exceeds the forecast probability by roughly 4--8
percentage points (for example 55\% observed in the 50\% bin, and 88\% in the 80\% bin), and the
10th--90th percentile band excludes the diagonal in each of those bins. Only in the 90\% bin,
where the band extends from about 90 to 96\%, is the departure consistent with sampling noise.
The underconfidence at 48\,h and 10\,mm is therefore a modest but real feature of the network,
though the control departs from the diagonal by a comparable amount in the opposite sense (it is
overconfident, with observed frequencies below the forecast probabilities), so the network is
still no worse calibrated than the control there.

\begin{figure}[H] \centering
\includegraphics[width=\textwidth]{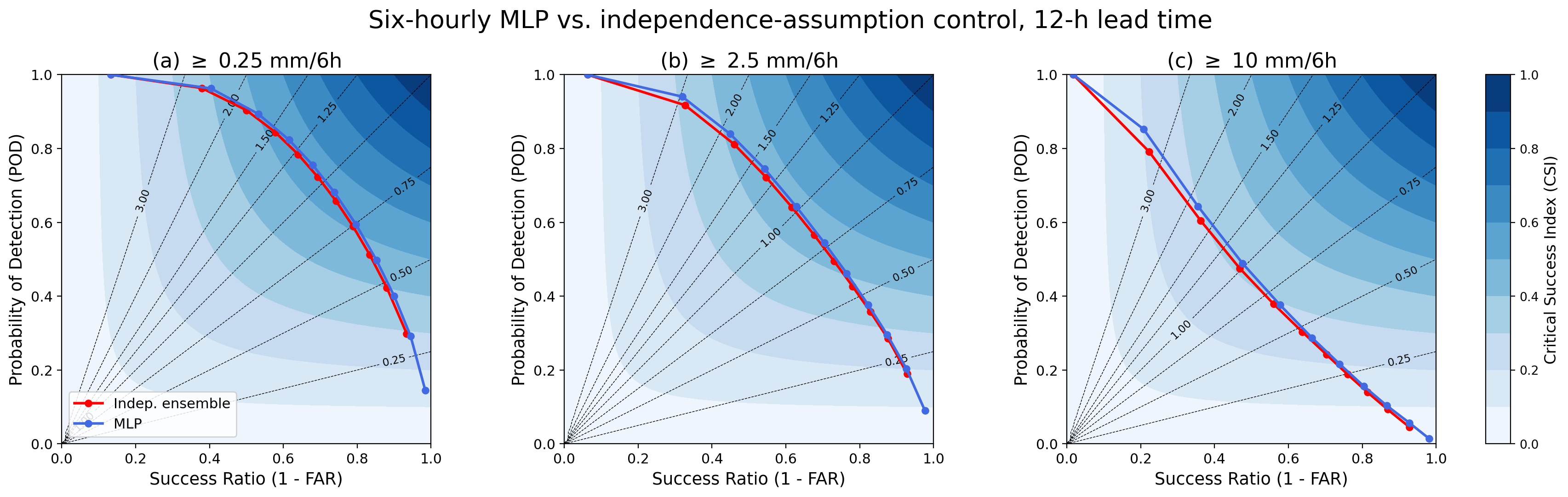} \caption{Performance
diagrams (probability of detection vs.\ success ratio, with shaded critical success index and
dashed frequency-bias contours) at 12-h lead time, comparing the six-hourly network (blue)
against the independence-assumption ensemble control (red). Panels show curves for six-hourly
accumulated precipitation exceeding (a) 0.25\,mm, (b) 2.5\,mm, and (c) 10.0\,mm, with each curve
built by cumulating contingency-table counts across forecast-probability bins from 0 to 100\% in
10\% increments.} \label{fig:perfdiag_12h} \end{figure}

\begin{figure}[H] \centering
\includegraphics[width=\textwidth]{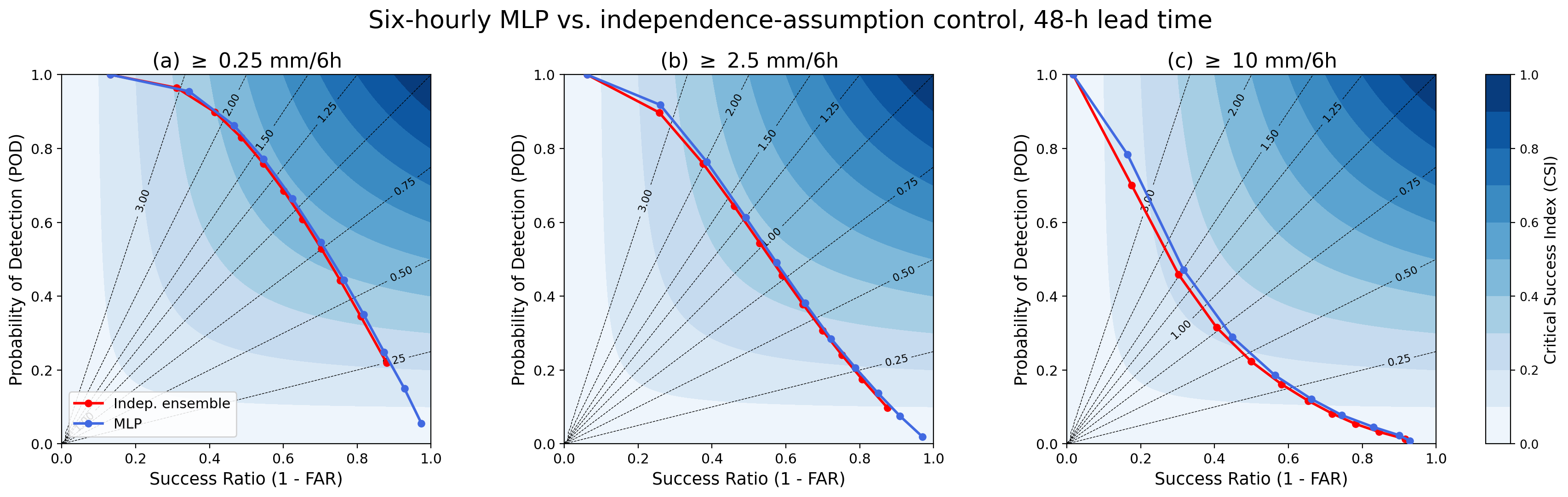} \caption{As in
Fig.~\ref{fig:perfdiag_12h}, but for 48-h lead time.} \label{fig:perfdiag_48h} \end{figure}

Performance diagrams \citep{roebber2009} provide a complementary view of forecast quality by
jointly displaying the probability of detection (POD) and success ratio (SR, equal to 1 minus the
false alarm ratio), with shaded background contours of the critical success index (CSI) and
dashed lines of constant frequency bias. Curves are constructed from the same per-bin contingency
tables underlying Figs.~\ref{fig:relia_12h} and~\ref{fig:relia_48h}, by treating each of the 11
saved forecast-probability bins (0, 10, 20, \ldots, 100\%) as a probability threshold above which
the forecast is counted as ``yes,'' and cumulating hit, false-alarm, and miss counts from that
bin through the highest bin. Figures~\ref{fig:perfdiag_12h} and~\ref{fig:perfdiag_48h} show these
diagrams for the 12-h and 48-h lead times, comparing the network against the
independence-assumption ensemble control at each of the three verification thresholds. Points at
the same nominal probability threshold do not correspond to equal numbers of ``yes'' forecasts
for the two methods, because their probabilities are calibrated differently; the two curves are
therefore best compared along their length and against the dashed bias lines rather than at
matching threshold labels. In contrast to the clear BSS separation reported above, the two curves
lie close together at both lead times and for all three thresholds, frequently overlapping. The
network's curve lies marginally on or above the control's over most of the probability range, and
its maximum CSI exceeds the control's by only 0.004 to 0.010 at every threshold and lead time
shown.

This is not a contradiction of the BSS results, but a consequence of what each diagnostic
measures. A performance diagram is built entirely from which forecast-probability bin a case is
placed in, so it reflects discrimination -- the ability to rank at-risk situations above
less-risky ones -- and is largely insensitive to whether the numeric probability value attached
to a bin is itself trustworthy. Because the control consumes exactly the same hourly marginal
distributions that are fed to the network as input features, the two methods rank cases very
similarly, and so their discrimination, and hence their performance-diagram curves, are nearly
identical. What differs sharply between the two methods is calibration: as shown above, the
control systematically overstates its own confidence at high-probability bins, while the
network's forecast probabilities track the diagonal closely over the probability range. BSS,
unlike a performance diagram, penalizes this miscalibration directly, which is why the network's
BSS advantage over the control is real and meaningful even though the two methods' performance
diagrams are nearly indistinguishable. In short, the network's demonstrated advantage over the
naive independence assumption lies mainly in probabilistic calibration rather than in event
discrimination.

\begin{figure}[H] \centering
\includegraphics[width=\textwidth]{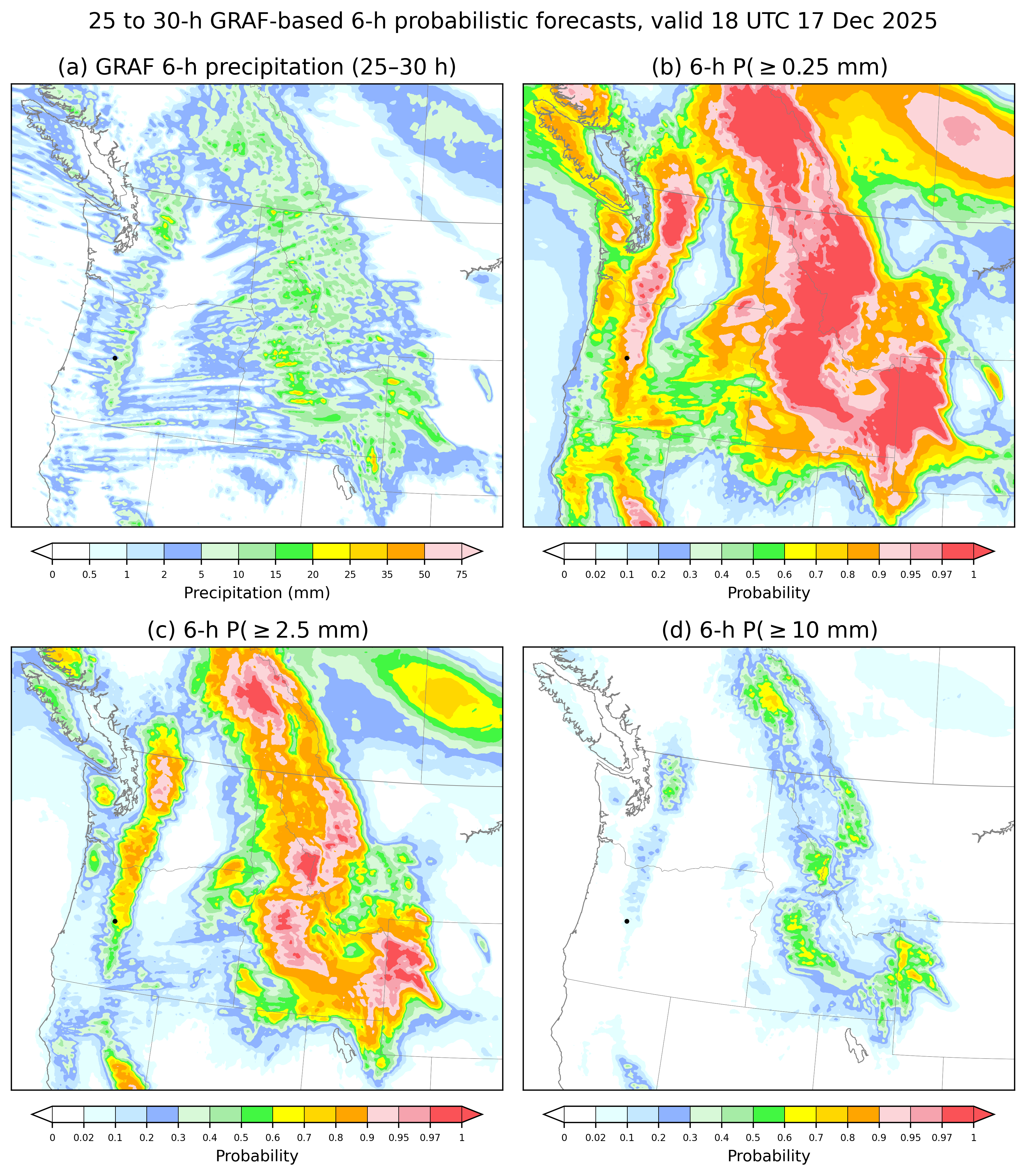}
\caption{Four-panel plot of six-hourly probabilistic forecasts for the initial condition of
12~UTC 16~December~2025 at 30-h lead time. (a) GRAF deterministic six-hourly precipitation (mm);
(b) probability of six-hourly accumulation $\geq$0.25\,mm; (c) probability $\geq$2.5\,mm; (d)
probability $\geq$10\,mm. The dot marks the location ($44^{\circ}$N, $122.25^{\circ}$W) examined
in Fig.~\ref{fig:sample_36h}.} \label{fig:4panel_24h} \end{figure}

\begin{figure}[H] \centering
\includegraphics[width=\textwidth]{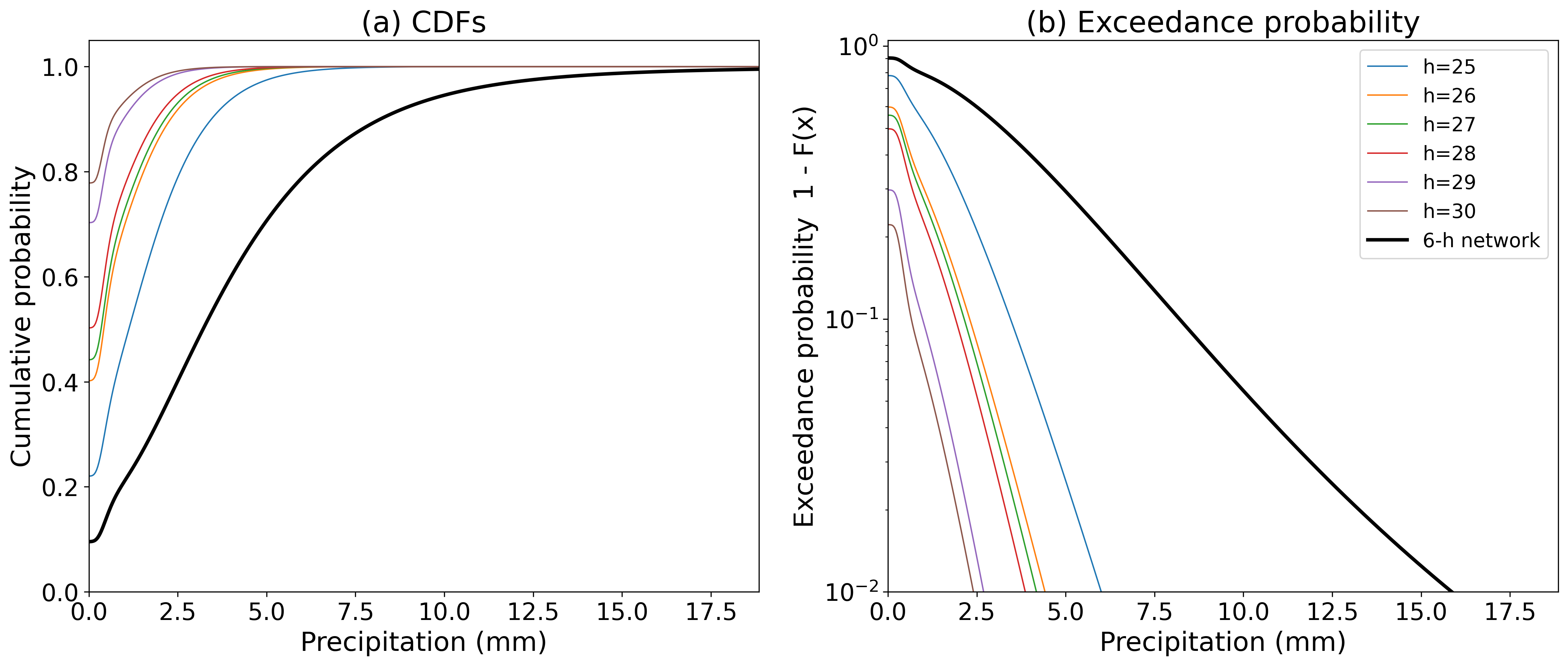}
\caption{Predicted CDF (left panel) and exceedance probability (right panel) at $44^{\circ}$N,
$122.25^{\circ}$W for the initial condition of 12~UTC 16~December~2025 at 30-h lead time. Thin
colored lines show the six consecutive hourly gamma-mixture distributions (lead hours 25--30);
the thick black line shows the six-hourly output distribution.} \label{fig:sample_36h}
\end{figure}

Figure~\ref{fig:4panel_24h} shows the spatial structure of the six-hourly probabilistic forecasts
for a representative Pacific Northwest US case from the out-of-sample test period. The initial
condition is 12 UTC 16 December 2025, the lead time is 30~h, and the forecast is for the six
hours ending 18 UTC 17 December 2025. The GRAF deterministic precipitation field (panel a) shows
widespread light precipitation (mostly under about 10\,mm) along the Cascades and Coast Ranges
and over the northern Rockies, and the probability fields (panels b--d) exhibit terrain-related
spatial structure that tracks the deterministic precipitation pattern, with the 10\,mm
probabilities (panel d) confined to the mountainous terrain. The dot marks the diagnostic
location ($44^{\circ}$N, $122.25^{\circ}$W) used in Fig.~\ref{fig:sample_36h}.

Figure~\ref{fig:sample_36h} shows the predicted cumulative distribution function (CDF) and
exceedance probability at the marked location for the same initial condition at 30-h lead time.
The six thin colored lines show the six consecutive hourly distributions entering the network
(lead hours 25--30), and the thick black line shows the six-hourly output distribution. The
network combines the information carried across the six individual hours into a single coherent
six-hourly predictive distribution. The probability of any precipitation during the window is
about 90\%, whereas the hourly probabilities range from about 22\% to 78\%, and the six-hourly
CDF is shifted well to the right of any single hourly CDF. The six-hourly probability of
exceeding 5\,mm is roughly 30\%, whereas no single hour has more than about a 5\% chance of
exceeding that amount; summed over six hours, even modest hourly rates produce a six-hourly total
well beyond what any individual hour's distribution would suggest.

\section{Summary} \label{sec:summary}

We have described a method for generating six-hourly probabilistic precipitation forecasts, for
lead times from 6 to 72\,h, from a sequence of six consecutive hourly probabilistic forecasts
produced by the Attention Residual U-Net. The core of the approach is a compact network, with
50{,}202 trainable parameters, in which a small GRU summarizes the six hourly gamma-mixture
parameter sets and statistics of the spatial texture of the raw forecast, and a fully connected
network combines this summary with seasonal, time-of-day, terrain and cross-hour coherence
predictors to produce the ten parameters of a six-hourly zero-inflated three-component Gamma
mixture. Several design choices proved important to this approach. Operating on distributional
parameters rather than raw fields avoids storing full-resolution spatial feature data for
training and allows the six-hourly model to leverage the calibration properties already embedded
in the hourly forecasts. Training uses importance sampling biased toward grid points with a
higher probability of precipitation and a larger forecast expected accumulation, with spatial
thinning to reduce redundancy among samples, combined with a stricter MRMS quality threshold ($Q
> 0.6$) than used for hourly training, so that the training set emphasizes meteorologically
relevant situations and limits contamination from poor-quality radar retrievals. The output layer
is initialized climatologically via an EM fit of a three-component Gamma mixture. Input features
are Z-score normalized, appropriate for the wide and \textit{a priori} unknown dynamic range of
the gamma-mixture parameters across lead times and seasons. Finally, the model is trained with an
NLL loss using log-sum-exp stabilization and per-sample capping, a strictly proper
scoring rule that rewards both the calibration and the sharpness of the predicted six-hourly
distribution.

The six-hourly network's Brier Skill Score is higher than that of a naive independence-assumption
ensemble control, which sums independent Monte Carlo draws from the same hourly gamma-mixture
forecasts, at all 12 lead times from 6 to 72\,h and for thresholds of 0.25, 2.5 and 10.0\,mm, and
a calendar-day bootstrap shows the difference to be statistically distinguishable from zero in
every case. The margin is largest and increases with lead time at the 0.25\,mm threshold, from
0.029 at 6\,h to 0.063 at 72\,h. Reliability diagrams show that the network is well calibrated,
while the control is consistently overconfident. Performance diagrams built from the same
contingency tables tell a different, complementary story: the two methods' curves lie close
together at both lead times shown, with maximum critical success index differing by at most
0.010, the network's being marginally higher. Together, these results indicate that the network's
demonstrated advantage over the naive independence assumption lies mainly in probabilistic
calibration -- the numeric trustworthiness of its forecast probabilities -- rather than in its
ability to discriminate events from non-events, since the two methods consume the same hourly
marginal information and therefore rank cases very similarly. An example case study showed a
realistic-looking relationship between precipitation amount and terrain relief.

Several aspects of the design should be flagged explicitly. The network's components -- the
texture and cross-hour coherence features, the GRU, the removal of dropout before the output
layer, and the third Gamma component -- were adopted after development experiments, chiefly at
the 12-h lead time, that compared reliability and BSS. Those comparisons were made on forecasts
from the same 2025 test period used here, so the reported skill may be subject to a modest
selection effect, and the quantitative effect of each component is not documented in this
article. Other choices were fixed at their initial values rather than chosen by systematic comparison: the
hidden-layer widths $[128, 144, 72, 36, 12]$ and the expand-then-compress shape, the GRU width of
48, and the two-pair cyclic predictors were not tuned in a controlled way, and no hyperparameter
search was performed. Using more than one verification year would be a more stringent test. As
described in Sec.~\ref{sec:seed}, results at 72\,h are for a favorable seed; the effect of seed
choice at other leads was not examined. Finally, incorporating additional predictors beyond those
used here---such as GFS large-scale moisture fields---could potentially improve the six-hourly
distributional estimates, particularly for thresholds associated with convective precipitation.

Unlike an ensemble forecast, a method like this does not provide ensemble scenarios, which may
be used, for example, in ensemble streamflow prediction \citep{schaake2007hepex}. This is an
admitted limitation of the method. \citet{bellier2023}
address a related problem -- deriving coherent multi-temporal-scale precipitation guidance from
finer-resolution ensemble forecasts -- using a modulation mechanism applied on top of standard
univariate calibration and reordering methods such as the Schaake shuffle; unlike the MLP
described here, that approach preserves ensemble scenarios by construction, at the cost of
depending on an underlying ensemble forecast rather than a single deterministic-model
probabilistic output.

A second admitted limitation concerns missing or degraded hourly input. The network requires all six
hourly gamma-mixture parameter sets spanning the accumulation window to be present; every script
in this pipeline that assembles the six-hour input vector, whether for verification (as used
here) or for the standalone case-study plots, checks for the existence of all six hourly files
and simply declines to produce a six-hourly forecast -- returning no result for that
initialization and lead time, rather than substituting a default, an interpolated, or a
climatological value -- if even one of the six is missing. This all-or-nothing behavior is
adequate for retrospective verification, where incomplete cases can simply be excluded, as was
done here, but it is not yet a complete solution for operational use. A single missing or
flagged-bad hourly input, for example from a GRAF or GFS data outage upstream of the hourly
Attention ResUNet, would presently cause the entire six-hourly forecast for that initialization and
lead time to be withheld. No fallback strategy, such as substituting a persisted forecast from an
earlier hour, an hourly climatology, or the mean of the remaining five hours, has been
implemented or evaluated, and this would need to be addressed before the method could be relied
upon in an operational setting where upstream data gaps do occur.

In future work, we expect to explore related methods for other accumulation periods such as 12 or
24 h, and also for shorter accumulation periods such as 0.25 h.

\section*{Acknowledgements}
The Weather Company's Monte Flora is thanked for a constructive, detailed review of this manuscript.



\end{document}